# Reliable Gene Mutation Prediction in Clear Cell Renal Cell Carcinoma through Multi-classifier Multi-objective Radiogenomics Model

Xi Chen[1], Zhiguo Zhou[2]*, Raquibul Hannan[2], Kimberly Thomas[3], Ivan Pedrosa[4], Payal Kapur[5,6], James Brugarolas[6,7], Xuanqin Mou[1] and Jing Wang[2]*

[1] School of Electronic and Information Engineering, Xi'an Jiaotong University, Xi'an, Shaanxi 710049, China

[2] Department of Radiation Oncology, University of Texas Southwestern Medical Center, Dallas, Texas 75235, USA

[3] Department of Radiation Oncology, Weill Cornell Medicine, New York, New York 10065, USA

[4] Department of Radiology, University of Texas Southwestern Medical Center, Dallas, Texas 75235, USA

[5] Department of Pathology, University of Texas Southwestern Medical Center, Dallas, Texas 75235, USA

[6] Kidney Cancer Program, Simmons Cancer Center, University of Texas Southwestern Medical Center, Dallas, Texas 75235, USA

[7] Department of Internal Medicine, University of Texas Southwestern Medical Center, Dallas, Texas 75235, USA

Email: Zhiguo.Zhou@UTSouthwestern.edu, Jing.Wang@utsouthwestern.edu

**Abstract:** Genetic studies have identified associations between gene mutations and clear cell renal cell carcinoma (ccRCC). Since the complete gene mutational landscape cannot be characterized through biopsy and sequencing assays for each patient, non-invasive tools are needed to determine the mutation status for tumors. Radiogenomics may be an attractive alternative tool to identify disease genomics by analyzing amounts of features extracted from medical images. Most current radiogenomics predictive models are built based on a single classifier and trained through a single objective. However, since many classifiers are available, selecting an optimal model is challenging. On the other hand, a single objective may not be a good measure to guide model training. We proposed a new multi-classifier multi-objective (MCMO) radiogenomics predictive model. To obtain more reliable prediction results, similarity-based sensitivity and specificity were defined and considered as the two objective functions simultaneously during training. To take advantage of different classifiers, the evidential reasoning (ER) approach was used for fusing the output of each classifier. Additionally, a new similarity-based multi-objective optimization algorithm (SMO) was developed for training the MCMO to predict ccRCC related gene mutations (*VHL*, *PBRM1* and *BAP1*) using quantitative CT features. Using the proposed MCMO model, we achieved a predictive area under the receiver operating characteristic curve (AUC) over 0.85 for *VHL*, *PBRM1* and *BAP1* genes with balanced sensitivity and specificity. Furthermore, MCMO outperformed all the individual classifiers, and yielded more reliable results than other optimization algorithms and commonly used fusion strategies.

## 1. Introduction

Kidney cancer, most predominantly, renal cell carcinoma (RCC) remains one of the most common renal malignancies with 63,990 new cases expected to be diagnosed and with 14,400 deaths in the United States in 2017 (Siegel *et al*, 2017). Clear cell RCC (ccRCC) is the most abundant (~75%) subtype of RCC and the most likely to metastasize outside the kidney (Motzer *et al*, 2002). Most cases of ccRCC present with somatic (or germline) inactivating mutations in the von Hippel–Lindau tumor suppressor (*VHL*) gene, which are generally absent in other cancers (Gnarra *et al*, 1994; Varela *et al*, 2011; Guo *et al*, 2012; Cancer Genome Atlas Research Network, 2013). Several other mutations in genes involved in regulating chromatin states, including those in the BRCA1-associated protein 1 (*BAP1*), polybromo 1 (*PBRM1*), SET domain containing 2 (*SETD2*), and lysine (K)-specific demethylase 5C (*KDM5C*), were recently identified (Dalgliesh *et al*, 2010; Varela *et al*, 2011; Duns *et al*, 2010; Peña-Llopis *et al*, 2012). Mutations in *BAP1* and *SETD2* were found to be associated with advanced stage and poor outcome (Cancer Genome Atlas Research Network, 2013; Hakimi *et al*, 2013; Kapur *et al*, 2013). The genes mutated within a tumor can be used as biomarkers and may help with prognosis, treatment selection, and treatment response prediction. However, inter- and intra-tumoral heterogeneity in gene mutations has previously been described in ccRCC (Gerlinger *et al*, 2012, 2014; McGranahan and Swanton, 2015). As ccRCC metastasizes, additional gene mutations accumulate. Because the complete gene mutational landscape is hard to be characterized for each patient through biopsy and sequencing assays, a non-invasive tool would be useful to identify the mutations within the tumor.

Radiogenomics (Rutman and Kuo, 2009; Jaffe, 2012; Kuo and Jamshidi, 2014; Karlo *et al*, 2014; Shinagare *et al*, 2015; Sala *et al*, 2017), an integrated approach that combines radiology and genomics, is based on extracting and analyzing amounts of data from medical images and clinical information by high-throughput computing. Therefore, radiogenomics is a promising solution for predicting gene mutation in ccRCC. Contrast-enhanced computed tomography (CT) is commonly used to diagnose and characterize renal masses, monitor growth in pathologically-proven RCC undergoing active surveillance, assess RCC location and extent, and determine stage and treatment response (Stewartmerrill *et al*, 2015; Motzer *et al*, 2017). Furthermore, the diagnostic standard of reference has expanded to the genomic level and has led to the attempt to use imaging as a noninvasive determinant of mutational status (Reznek, 2004; Powles and Albers, 2012; Carles *et al*, 2012; Kuo and Yamamoto, 2011). Therefore, a CT based radiogeneomics predictive model would be helpful.

In recent years, researchers have investigated predictable and systematic associations between imaging features and underlying molecular and genomic alterations in different cancers. Yamamoto *et al* (2012) carried out a radiogenomic analysis of breast cancer with MRI, a novel approach that may help reveal the underlying molecular biology of breast cancers. Gevaert *et al* (2017) used CT image features to predict the mutation status of EGFR in non–small cell lung cancer (NSCLC). Aerts *et al* (2014) revealed that a prognostic radiomic feature set, capturing intra-tumor heterogeneity, is associated with underlying gene-expression patterns. One study reported associations between CT features of 58 ccRCCs and the underlying karyotype (Sauk *et al*, 2011). Others studied the association between CT imaging features and mutational status of ccRCC (Karlo *et al*, 2014; Shinagare *et al*, 2015). For example, the *BAP1* mutation was associated with ill-defined tumor margins and calcification (Shinagare *et al*, 2015).

By quantitatively analyzing large amounts of information from medical images, radiogenomics holds great potential to predict gene mutation. However, several challenges need to be addressed to build an optimal predictive model. First, a single classifier is typically used to build a radiogenomics predictive model. Aerts *et al* (2014) used the Cox proportional hazards regression model to predict survival in patients with lung and head-and-neck cancer. Other researchers tested different types of classifiers and chose one or two "preferred" ones for specific applications. Valdes *et al* (2016) evaluated three different classifiers, including decision trees, random forests, and RUSBoost, to predict radiation pneumonitis in patients with stage I NSCLC treated with stereotactic body radiation therapy (SBRT). Higher accuracy was achieved when the RUSBoost algorithm was used with regularization. These findings indicate how difficult it is to select a "preferred" classifier for a specific application. Instead of trying to find the most suitable classifier for a particular application, a model that combines multiple classifiers can fully use information from different classifiers to improve accuracy in radiogenomics. Second, most current radiogenomics models adopt a single objective function (e.g. accuracy, AUC), which may not be a good measure for building the predictive model, especially when positive and negative cases are imbalanced. To overcome the disadvantages of using a single classifier and a single objective function, we sought to develop a multi-classifier multi-objective (MCMO) radiogenomics model predict most mutations in most commonly mutated genes in ccRCC. In MCMO, multiple classifiers are used for building the model and a multi-objective optimization algorithm is used for training the model.

## 2. Materials and Methods

### 2.1 Data

*2.1.1 Patients.* We conducted an institutional review board-approved, Health Insurance Portability and Accountability Act-compliant (HIPAA), retrospective study including 57 ccRCC patients from two independent cohorts. The first cohort consisted of 33 patients (median age 62 years, range 28–83) from the University of Texas Southwestern Medical Center (UTSW). The other cohort consisted of 24 patients (median age 59 years, range 26–74) from The Cancer Genome Atlas Kidney Renal Clear Cell Carcinoma (TCGA-KIRC) data collection. The TCGA-KIRC data collection is part of The Cancer Genome Atlas (TCGA), an ongoing project funded by the National Cancer Institute (NCI) and the National Human Genome Research Institute (NHGRI), which created an atlas of genetic changes related to more than 20 tumor types, including ccRCC. Clinical, genetic, and pathological data reside in the TCGA data portal, and radiological data is stored in The Cancer Imaging Archive (TCIA). Both TCGA and TCIA are accessible for public download (Smith *et al*, 2016; Clark *et al*, 2013).

All 57 patients fulfilled the following criteria: (a) histopathologic diagnosis of ccRCC and exome sequencing, including information on *VHL*, *PBRM1*, and *BAP1* gene mutations, considered as frequent mutations in ccRCC; (b) availability of images from a pretreatment contrast-enhanced CT including a corticomedullary phase. For each gene, the numbers of patients with mutation and without mutation are listed (table 1).

*2.1.2 CT Image features.* CT images from UTSW were acquired by GE LightSpeed VCT (GE Healthcare, Waukesha, WI) or TOSHIBA Aquilion ONE (Canon Medical Systems USA, Tustin, CA). CT image size was 512 × 512 with a pixel size of 0.7~0.9 mm, and slice thickness was 3 or 5mm.

**Table 1.** Number of patients

|  | VHL | | PBRM1 | | BAP1 | |
|---|---|---|---|---|---|---|
|  | Mutation | Non-mutation | Mutation | Non-mutation | Mutation | Non-mutation |
| UTSW | 26 | 7 | 19 | 14 | 5 | 28 |
| TCGA | 10 | 14 | 3 | 21 | 2 | 22 |
| Total | 36 | 21 | 22 | 35 | 7 | 50 |

CT images from TCIA were acquired by the SIEMENS Sensation 64 / Definition AS+ (Siemens Medical Solution, Malvern, PA), Philips Brilliance 64 (Philips Healthcare, Andover, MA), or GE LightSpeed VCT. CT image size was 512 × 512 with a pixel size of 0.7~1 mm and slice thickness was 1.25 or 5 mm.

The primary tumor contour was delineated by a radiation oncologist with 4 years of experience and reviewed by a radiation oncologist with 9 years of experience. Contrast enhanced CT images acquired during the corticomedullary phase were used in all 57 cases for image analysis. A region of interest (ROI) was drawn along the outer contour of the mass using the Velocity 3.2.0 software excluding adjacent tissues (e.g. renal parenchyma, peri-renal fat) (figure 1).

We resampled all images of the same slice thickness at 5mm. We only considered primary tumors and defined 43 quantitative image features describing tumor characteristics, including 13 geometry features, 9 intensity features, and 21 texture features (table 2). Geometry features describing tumor shape and size were calculated according to the actual pixel size. Features Size_X, Size_Y and Size_Z (table 2) describe the tumor size along the X, Y, and Z axes of the digital imaging and communications in medicine (DICOM) coordinate system (figure 2 (a)). The shape and location of the tumors differed from patient to patient. To intuitively describe the size of the tumor, a principal component analysis (PCA) was applied to the tumor contour points to transform the data into a new coordinate system. Size_P1 is the maximum 3 dimensional diameter of the tumor, measured as the largest pairwise Euclidean distance between the voxels on the surface of the tumor volume; size_P2 and size_P3 are the tumor size along the directions orthogonal to the direction of maximum size (figure 2(b)).

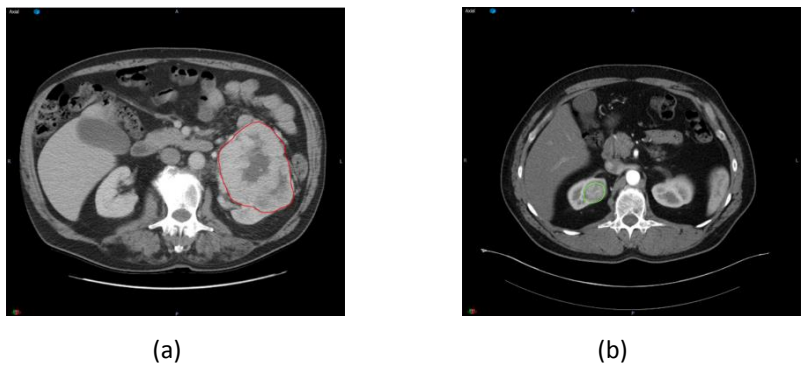

(a)　　　　　　　　　　(b)

**Figure 1.** Tumor contour of ccRCC investigated in this study. (a) one case from UTSW. (b) one case from TCIA.

Table 2. Quantitative CT Image Features

| Geometry features | Intensity features | Texture features |
|---|---|---|
| Volume | Minimum | Auto correlation |
| Size_X | Maximum | Contrast |
| Size_Y | Mean | Correlation |
| Size_Z | Median | Cluster prominence |
| Size_P1 | Sum | Cluster shade |
| Size_P2 | Variance | Dissimilarity |
| Size_P3 | Standard deviation | Energy |
| Roundness | Skewness | Entropy |
| Surface area | Kurtosis | Homogeneity |
| Compactness_1 | | Maximum probability |
| Compactness_2 | | Variance |
| Spherical disproportion | | Sum average |
| Surface to volume ratio | | Sum variance |
| | | Sum entropy |
| | | Difference variance |
| | | Difference entropy |
| | | Information measure of correlation_1 |
| | | Information measure of correlation_2 |
| | | Inverse difference |
| | | Inverse difference normalized |
| | | Inverse difference moment normalized |

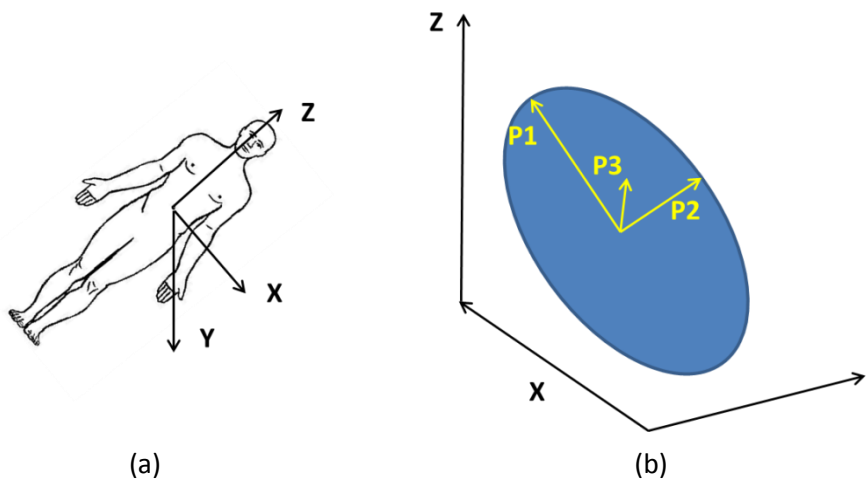

(a)  (b)

**Figure 2.** Coordinate systems. (a) The digital imaging and communications in medicine (DICOM) coordinate system. (b) Illustration of the transformation of the DICOM coordinate system into the PCA coordinate system.

Intensity features are first-order statistics that describe the distribution of the voxel intensities within the tumor on the CT image through commonly used basic metrics (Aerts *et al*, 2014). Intensity features provide information related to the gray-level distribution of the image; however, they do not provide any information on the relative position of the various gray levels over the image. Therefore, we included textural features that either describe the patterns or the spatial distribution of voxel intensities, which were calculated from the gray level co-occurrence matrix (GLCM) (Haralick *et al*, 1973). Texture matrix representation requires the voxel intensity values within the volume of interest to be discretized. In this work, voxel intensities were resampled into 64 equally spaced bins using a bin-width of 25 Hounsfield Units (HU). The detailed methodology of extracting intensity and texture features was previously described by Aerts *et al* (2014) (supplementary document). The calculated features of the 57 patients were normalized using the Z-scores method (Cheadle *et al*, 2003).

### 2.2. MCMO predictive model

*2.2.1 Evidential reasoning based classifier fusion* The evidential reasoning (ER) approach was used for fusing the individual classifier probability output (Yang and Xu, 2002, 2013). Our study is a binary classification problem (mutation or non-mutation). Assuming there are $M$ classifiers, for a test sample, the output probability of each classifier is denoted by $P_i = \{P_i^1, P_i^2\}, i = 1, \cdots, M$, which satisfies:

$$P_i^1 + P_i^2 = 1. \tag{1}$$

Where $P_i^1$ is output probability of gene mutation and $P_i^2$ is output probability of non-mutation. Assume the relative weight of each classifier as $\mathbf{w} = \{w_1, w_2, \cdots, w_M\}$, which satisfies the following constraint:

$$\sum_{i=1}^{M} w_i = 1, 0 \leq w_i \leq 1. \tag{2}$$

The final output probabilities $P_{fin}^j, j = 1,2$ are obtained by classifier fusion through the ER approach (Yang and Xu, 2002, 2013):

$$P_{fin}^j = ER(P_i^j, w_i), i = 1, \cdots, M, \ j = 1,2, \tag{3}$$

where $ER$ represents the ER analytic algorithm (Wang *et al*, 2006), which is calculated as:

$$P_{fin}^j = \frac{\mu \times \left[\prod_{i=1}^{M}(w_i P_i^j + 1 - \omega_i) - \prod_{i=1}^{M}(1-w_i)\right]}{1 - \mu \times \left[\prod_{i=1}^{M}(1-w_i)\right]}, \ j = 1,2, \tag{4}$$

where $\mu$ is calculated as:

$$\mu = \left[\sum_{j=1}^{2} \prod_{i=1}^{M}\left(w_i P_i^j + 1 - w_i\right) - \prod_{i=1}^{M}(1-w_i)\right]^{-1}. \tag{5}$$

For a binary classification problem, if $P_{fin}^1 > P_{fin}^2$, the test sample belongs to class 1; if $P_{fin}^1 < P_{fin}^2$, the test sample belongs to class 2; if $P_{fin}^1 = P_{fin}^2$, the test sample belong to either class. In this study, We sought to predict either the presence or absence of gene mutation. Therefore, if $P_{fin}^1 > P_{fin}^2$, we considered the test sample had mutation; if $P_{fin}^1 \leq P_{fin}^2$, we considered the test sample had non-mutation.

*2.2.2 Reliable outcome prediction based on output probability similarity* To obtain more reliable predictive results, reliable outcome prediction (RCP) is proposed and defined as maximize the similarity between predicted output probability and true label vector. For example, we assume two models that predict the *VHL* gene mutation with the label vector [1, 0]. Model A has prediction probabilities (0.8, 0.2) (the probability of mutation is 0.8, the probability of non-mutation is 0.2, and the threshold is 0.5), and model B has prediction probabilities (0.55, 0.45). As 0.8 is closer to 1 than 0.55, the result of model A is more reliable than that of model B. In other words, the similarity between the probability output of model A and the label vector is higher than those observed for model B, which means model A is more reliable than model B in this prediction.

In RCP, the aim is to maximize the similarity between predicted output probability and true label vector $T$ while training the single classifier model and weights. For a training sample, its label vector is denoted by $\boldsymbol{T} = [T_1, T_2]$. $T$ is a binary vector, $\boldsymbol{T} = [1,0]$ (mutation) or $\boldsymbol{T} = [0,1]$ (non-mutation). Assuming that the predictive model has $q$ parameters denoted by $\boldsymbol{R} = \{R_1, R_2, \cdots, R_q\}$, the objective function is expressed as:

$$f = \max_{\boldsymbol{W}, \boldsymbol{R}} \sum_{k=1}^{K} sim(P^k, T^k), \tag{6}$$

where $K$ represents the number of training samples and $sim$ is the similarity measure. Since the above problem can be considered as the similarity of probability distribution and the dice coefficient (Sung-Hyuk, 2007) is effective for measuring similarity, it is used here:

$$sim(P^k, T^k) = \frac{2\sum_{j=1}^{2} P_j^k T_j^k}{\sum_{j=1}^{2} (P_j^k)^2 + \sum_{j=1}^{2} (T_j^k)^2}. \tag{7}$$

Since a single objective may not be a good measure when the training dataset is imbalanced, we consider sensitivity and specificity simultaneously as a better solution as follows:

$$f_{sen} = \frac{TP}{TP+FN}, \quad f_{spe} = \frac{TN}{TN+FP}, \tag{8}$$

where $TP$ is the number of true positives, $TN$ is the number of true negatives, $FP$ is the number of false positives, and $FN$ is the number of false negatives. In our previous study, $f_{sen}$ and $f_{spe}$ were both considered as objective functions (Zhou et al, 2017).

However, $f_{sen}$ and $f_{spe}$ are label based measures, while we sought to maximize the similarity of probability output and the true label vector. Therefore, we defined the new similarity-based sensitivity and specificity denoted by $f_{sim\_sen}$ and $f_{sim\_spe}$, respectively. Assume that $\{P_{tp}^1, P_{tp}^2, \cdots, P_{tp}^{TP}\}$ represents the probability output of true positives and the corresponding true label vector is $\{T_{tp}^1, T_{tp}^2, \cdots, T_{tp}^{TP}\}$. The similarity of true positives $TP_{sim}$ is defined as:

$$TP_{sim} = \sum_{k=1}^{TP} sim(P_{tp}^k, T_{tp}^k) = \sum_{k=1}^{TP} \frac{2\sum_{j=1}^{2} P_{tp,j}^k T_{tp,j}^k}{\sum_{j=1}^{2} (P_{tp,j}^k)^2 + \sum_{j=1}^{2} (T_{tp,j}^k)^2}, \tag{9}$$

In gene mutation prediction, $j=1$ represents mutation, $j=2$ represents non-mutation, and $P^k$ is the mutation probability. Also, $P_{tp,1}^k = P^k$, $P_{tp,2}^k = 1 - P^k$, $T_{tp,1}^k = 1$, and $T_{tp,2}^k = 0$. Therefore, equation (9) can be simplified as:

$$TP_{sim} = \sum_{k=1}^{TP} \frac{P^k}{(P^k)^2 - P^k + 1}, \tag{10}$$

Similarly, we define the similarity of true negatives $TN_{sim}$, false positives $FP_{sim}$, and false negatives $FN_{sim}$:

$$TN_{sim} = \sum_{k=1}^{TN} \frac{1-P^k}{(P^k)^2 - P^k + 1}, \tag{11}$$

$$FP_{sim} = \sum_{k=1}^{FP} \frac{1-p^k}{(P^k)^2 - P^k + 1}, \tag{12}$$

$$FN_{sim} = \sum_{k=1}^{FN} \frac{p^k}{(p^k)^2 - p^k + 1}. \tag{13}$$

Then, $f_{sim\_sen}$, $f_{sim\_spe}$ are calculated as:

$$f_{sim\_sen} = \frac{TP_{sim}}{TP_{sim}+FN_{sim}}, \quad f_{sim\_spe} = \frac{TN_{sim}}{TN_{sim}+FP_{sim}}, \quad (14)$$

Our aim was to maximize the two similarity-based objective functions simultaneously as:

$$f_{sim} = \max_{\boldsymbol{w},\boldsymbol{R}}(f_{sim\_sen}, f_{sim\_spe}). \quad (15)$$

Once training is finished, the Pareto-optimal solution set is generated, and the best model parameters and weights are selected based on the clinical needs. In the following subsection, we describe a new algorithm that was developed to solve the similarity-based multi-objective optimization problem.

### 2.3. Similarity-based multi-objective optimization (SMO) algorithm

Multi-objective evolutionary algorithms (MOEA) have demonstrated the superior performance for multi-objective optimization (Deb, 2001). Based on MOEA, we have proposed an iterative multi-objective immune algorithm (IMIA), which adopts the traditional sensitivity and specificity as the optimized objective functions (Zhou et al, 2017). Based on IMIA, we propose a new SMO algorithm. The major difference between IMIA and SMO is that the reliability is measured through similarity between output probability and label vector during the training process. Additionally, weighting coefficients needs to be optimized in SMO. For conciseness, we just gave a brief description of SMO and focused on the difference between SMO and IMIA. For a full and detailed algorithm, please refer to our previous paper (Zhou et al, 2017).

As IMIA, SMO consists of the 7 steps: initialization, cloning, mutation, deletion, solution updating, termination and best solution selection. In initialization, model parameters $\boldsymbol{R}$ and weights $\boldsymbol{w}$ were both initialized. We generated the initial solution set $D(t) = \{d_1,\cdots,d_H\}$ ($t = 0$) randomly, $d_i(i = 1,\cdots,H)$ is a particular solution, $H$ is the number of solutions, and $t$ is the number of generation. In cloning step, there is a big difference. In SMO, new similarity-based proportional cloning operation was proposed, where the solution with higher similarity was reproduced multiple times. Specifically, the clonal time $CLT_i$ for each solution is calculated as:

$$CLT_i = \left\lceil n_c \times \frac{sim(d_i)}{\sum_{i=1}^{H} sim(d_i)} \right\rceil, \quad (15)$$

where $n_c$ is the expected value of the clonal solution set and $\lceil \ \rceil$ is the ceiling operator. The similarity measure for solution $d_i$ denoted by $sim(d_i)$ is calculated as:

$$sim(d_i) = \sum_{k=1}^{K} \frac{2\sum_{j=1}^{2} P_j^k T_j^k}{\sum_{j=1}^{2}\left(P_j^k\right)^2 + \sum_{j=1}^{2}\left(T_j^k\right)^2}, \quad (16)$$

where $K$ is the number of training samples and $T_j^k$ is the label vector.

The mutation and deletion in SMO are the same as those in IMIA. In this paragraph, "mutation" refers to the operation performed on the cloned solution set, not the gene mutation. The mutation probability threshold $MP$ is determined empirically and an operation probability $RP_i$ is generated randomly. If $RP_i > MP$, a mutation operation is performed in which a new solution $d_i^m$ was generated randomly and replace the original solution $d_i$. Then, a newly generated mutated solution set $M(t)$ and solution set $D(t)$ constitute the new solution set denoted by $F(t)$. Same solutions in $F(t)$ were removed and a new solution set $DF(t)$ is generated.

In solution updating step of SMO, the aim is to select $H$ solutions from $DF(t)$ to maintain the population size. For each solution, we can obtain the similarities based on the probability outputs of all the training samples, according to equation (16). $f_{sim\_sen}$ and $f_{sim\_spe}$ can also be calculated according to equation (14). Using the MOEA (Deb, 2001; Deb *et al*, 2002), we selected $H$ solutions. Unlike most traditional MOEAs, the solution in $DF(t)$ is sorted according to the similarity of each solution. Then, the new solution set $UD(t)$ is generated.

When *t* reaches the maximal number of generations $G_{max}$, the algorithm terminates. The best solution is selected from the Pareto-optimal solution set $UD(G_{max})$ according to clinical needs. We selected the best solution according to the similarity-based sensitivity, specificity, and AUC. First, the thresholds $T_{sim\_sen}$ and $T_{sim\_spe}$ are determined for similarity-based sensitivity and specificity according to clinical needs. Second, for each solution $d_i$ in $UD(G_{max})$, we calculate its similarity-based sensitivity $f_{sim\_sen}^i$ and specificity $f_{sim\_spe}^i$. If $f_{sim\_sen}^i > T_{sim\_sen}$ and $f_{sim\_spe}^i > T_{sim\_spe}$, $d_i$ is selected as a candidate solution. Third, we select the solution with the highest AUC from the candidate solutions as our final solution.

### 2.4 Training and testing procedure of the MCMO model

The training process mainly consists of three stages: feature calculation, feature selection, and predictive model construction. To achieve optimal performance for each classifier, we adopted our multi-objective feature selection method (Zhou *et al*, 2017). After selecting the features for each classifier, model parameters $R_i$ and weights $w_i$ ($i = 1,2,...M$) were trained. The workflow is illustrated in figure 3.

The testing process consists of three stages (figure 4). For a test sample, first, the features for each classifier are selected; second, each classifier outputs a probability $P_i^j$ ($j = 1,2; i = 1,2,...M$); third, the final mutation probability $P_{fin}^j$ is obtained by combining all $P_i^j$ and $w_i$ using the ER approach. Then the label can be determined.

We used six different classifiers in the MCMO model, including support vector machine (SVM) (Keerthi and Lin, 2003), logistic regression (LR) (Freedman, 2009), discriminant analysis (DA) (Hastie and Tibshirani, 1996), decision tree (DT) (Breiman, 2001), K-nearest-neighbor (KNN) (Keller *et al*, 2012), and naive Bayesian (NB) (Goldszmidt and Moises, 1997). Since SVM has two model parameters and other classifiers use default parameters, we train eight parameters $\{R_{SVM\_1}, R_{SVM\_2}, w_1, ... w_6\}$ for the predictive model.

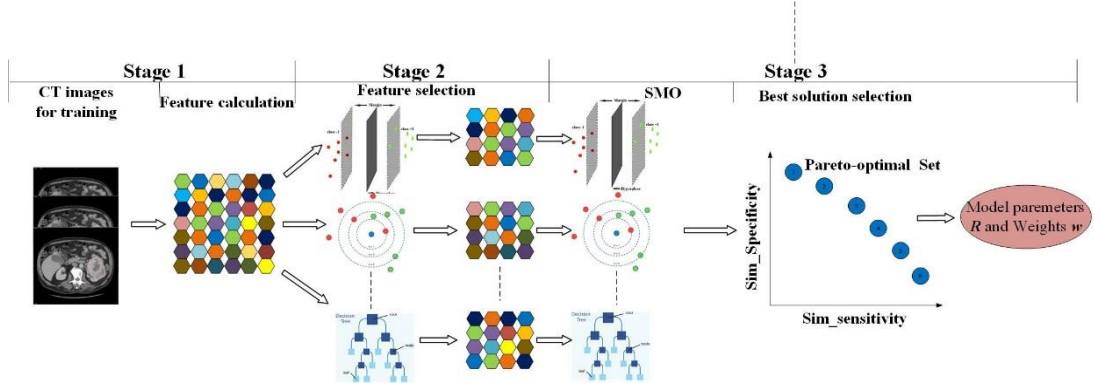

**Figure 3.** Training process of MCMO predictive model.

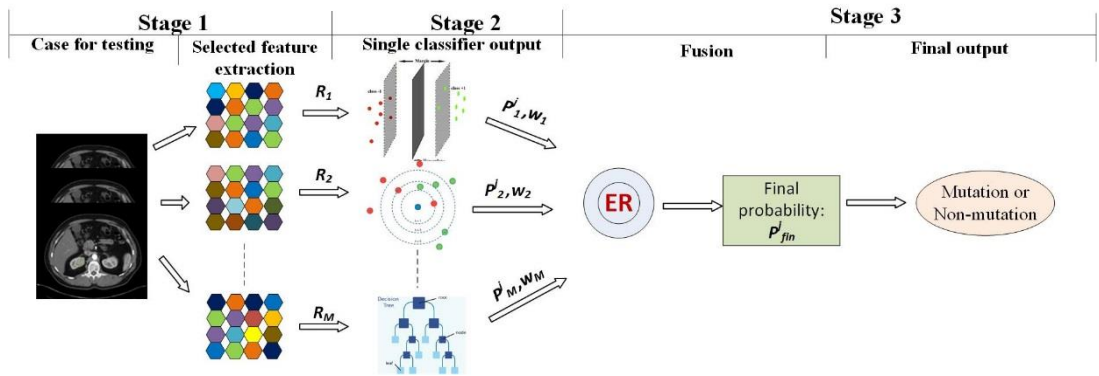

**Figure 4.** Testing process using MCMO predictive model.

## 3. Results

### *3.1 Experimental setup*

In MCMO, the population number *H* and the maximal generation number $G_{max}$ were both set to 100. In the clone operation, $n_c$ was set to 200. In the mutation operation, the mutation probability *MP* was set to 0.9. The proposed MCMO predictive model was compared to three types of predictive model: (1) models with single classifiers; (2) models with different multi-objective optimization; (3) models with different fusion strategies. Because multi-objective optimization is better than the single-objective model (Zhou *et al*, 2017), we did not compare our proposed model to models with single-objective optimization. Area under the receiver operating characteristic curve (AUC), accuracy, sensitivity, specificity, similarity-based sensitivity (Sim-sensitivity), and similarity-based specificity (Sim-specificity) were used for evaluating the model performance.

In our study, eight parameters in the predictive model need to be trained and our dataset has 57 cases. We adopted two-fold cross-validation in this work. Cross-validation is a widely used model validation technique which can test the model's ability to predict new data that were not used in training, in order to flag problems like overfitting (Kohavi, 1995). One round of cross-validation involves partitioning a sample of data into complementary subsets. In our work, two-fold cross-validation was used, where for each round, half cases (training set) were selected randomly for training and the other half cases (validation set) were used for validation, then reverse. To reduce variability caused by subset partition, ten rounds of two-fold cross-validation were performed for each model, and the validation results were averaged over the 10 rounds to give an estimate of the model's predictive performance (table 3). Prediction results of AUC, accuracy, sensitivity and specificity of training set were also listed in table 3. The prediction accuracy of training set is higher than that obtained from the validation set, which is considered as normal for a machine learning algorithm. On the other hand, the accuracies of the training set are not close to 1 and the predictive results on the validation set are acceptable. In the remaining of the paper, all the prediction results were from the validation set.

Table 3. Results of MCMO prediction models (training set vs. validation set)

| Gene | Data set | AUC | Accuracy | Sensitivity | Specificity |
|---|---|---|---|---|---|
| VHL | Training set | 0.96±0.02 | 0.91±0.01 | 0.93±0.02 | 0.88±0.02 |
| | Validation set | 0.88±0.01 | 0.81±0.02 | 0.79±0.04 | 0.86±0.02 |
| PBRM1 | Training set | 0.95±0.01 | 0.90±0.03 | 0.87±0.03 | 0.92±0.02 |
| | Validation set | 0.86±0.02 | 0.78±0.02 | 0.75±0.02 | 0.80±0.02 |
| BAP1 | Training set | 0.97±0.01 | 0.92±0.01 | 0.9±0.03 | 0.92±0.01 |
| | Validation set | 0.93±0.02 | 0.90±0.02 | 0.87±0.02 | 0.90±0.03 |

## 3.2 Selected features for the three genes

We obtained the classifier-specific feature subset and summarized the features selected by all (six) classifiers (table 4). Statistics using the unpaired *T* test were also listed. The feature selected by all classifiers indicates that the feature is important to predict mutation. *P*-value smaller than 0.05 indicated significant differences between presence and absence of mutations. However, four selected features (Minimum, Contrast, Variance and Sum variance) for the *PBRM1* gene and three selected features (Variance, Sum average and Sum variance) for the *BAP1* gene had *P*-values greater than 0.05. Because we selected the optimal features according to multi-objective model, using the AUC as the figure of metric, these selected features did not necessarily have *P*-values smaller than 0.05. Example boxplots were plotted to show the potential of a single feature for differentiating between the presence and absence of mutations (figure 5).

Table 4. Selected features by all (six) classifiers for *VHL*, *PBRM1* and *BAP1* genes.

| | Geometry (*P*-value) | Intensity (*P*-value) | Texture (*P*-value) |
|---|---|---|---|
| VHL | **Mean (0.020)** | **Kurtosis (0.002)** | |
| PBRM1 | | Minimum (0.088) | Contrast (0.085) |
| | | **Mean (0.020)** | **Maximum probability (0.030)** |
| | | **Median (0.041)** | Variance (0.162) |
| | | **Skewness (0.042)** | Sum variance (0.168) |
| | | **Kurtosis (0.005)** | |
| BAP1 | **Size_X (0.006)** | **Sum (0.048)** | **Homogeneity (0.023)** |
| | | | Variance (0.171) |
| | | | Sum average (0.159) |
| | | | Sum variance (0.168) |

Two intensity features including Mean and Kurtosis are the most frequently selected features in *VHL* gene prediction. A histogram with a more elongated tail indicates smaller Kurtosis. A tumor with smaller Kurtosis is more likely to carry a *VHL* mutation (figure 5(a) and figure 6). For the *PBRM1* gene prediction, nine features from intensity and texture features, were most frequently selected. A boxplot of the Mean is illustrated in figure 5(b). Five features were selected as the most prominent contributors in *BAP1*. Four texture features were selected as follows: Homogeneity, Variance, Sum average, and Sum variance. A lower similarity in intensity between a voxel and its neighbors led to higher Variance and Sum variance. A less uniform or more focal intensity distribution led to reduced Homogeneity. Therefore, larger Variance, and smaller Homogeneity were associated with the likelihood that a tumor carried a *BAP1* mutation. Boxplot of Homogeneity is illustrated in figure 5(c).

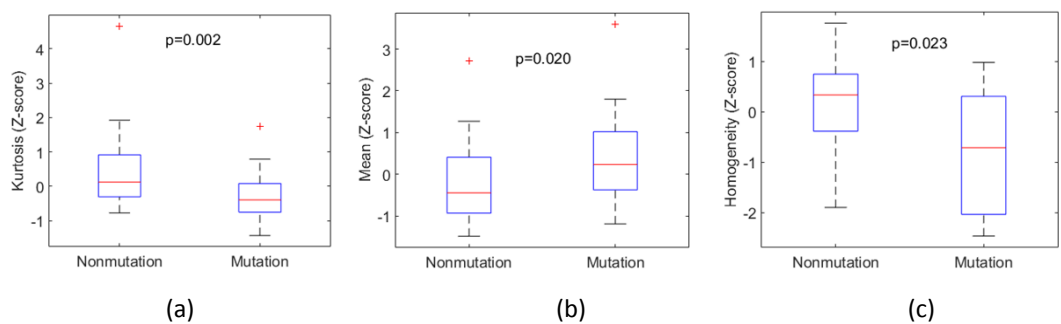

**Figure 5.** Boxplots for (a) Kurtosis (*VHL*), (b) Mean (*PBRM1*), and (c) Homogeneity (*BAP1*).

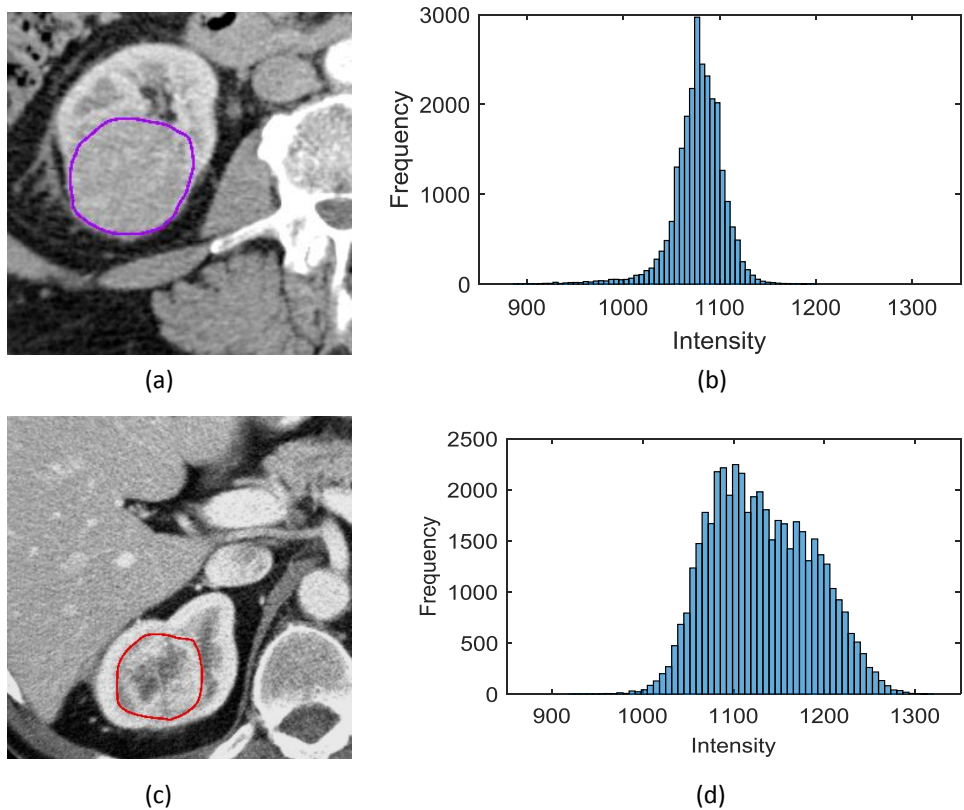

**Figure 6**. Kurtosis in tumor CT images. (a) A tumor without *VHL* mutation, (b) Histogram, with kurtosis= 26.57 (Z-score= 1.92), (c) A tumor with *VHL* mutation, (d) Histogram with kurtosis= 7.17 (Z-score= -1.33).

It is noted that one single feature may not achieve accurate predictive results. For each classifier, a feature set is necessary. For KNN classifier, a feature set consisting of 12 features (Volume, Size_P3, Minimum, Maximum, Mean, Sum, Variance, Standard deviation, Kurtosis, Cluster shade, Energy, Inverse difference) was selected to predict *VHL* mutation; while for LR classifier, a feature set consisting of 19 features (Volume, Size_Z, Size_P2, Roundness, Surface area, Mean, Standard deviation, Skewness, Kurtosis, Contrast, Dissimilarity, Energy, Entropy, Homogeneity, Sum entropy, Information measure of correlation_1, Information measure of correlation_2, Inverse difference, Inverse difference normalized) was selected to predict *VHL* mutation.

### *3.3 Performance evaluation of MCMO vs. single classifiers*

MCMO yielded better AUC, accuracy, sensitivity, and specificity results than other single classifiers (table 5). The prediction accuracy of MCMO is 0.81, 0.78, and 0.90 for *VHL*, *PBRM1*, and *BAP1* genes, respectively, with AUC >= 0.86 sensitivity > =0.75 and specificity > =0.80. MCMO yielded better results than other single classifiers. KNN is the best single classifier for the *VHL* and *PBRM1* genes, with specificities of 0.66 and 0.62, respectively. MCMO can achieve specificities of 0.86 and 0.80 for *VHL* and *PBRM1* genes, respectively. SVM and DA achieved similar results for the *BAP1* gene, which are better than other single classifiers, but sensitivities were only 0.57 and 0.63, respectively; the sensitivity obtained by MCMO was 0.87. Some single classifiers achieved higher sensitivities for the *VHL* gene, but the corresponding specificities were poor. MCMO achieved the highest AUC and accuracy with balanced sensitivities and specificities (difference < 0.1). Also, MCMO is a stable predictive model because its standard deviations are much smaller than those of the single classifiers.

### *3.4 Comparative study of objective functions*

One group of the Pareto-optimal solution set and the selected final solution in SMO is shown in figure 7. As described in section 2.3, the best solution was selected according to the similarity-based sensitivity, specificity (equation (14)), and AUC. First, thresholds $T_{sim\_sen}$ and $T_{sim\_spe}$ were determined for similarity-based sensitivity and specificity based on clinical needs. In this study, the thresholds $T_{sim\_sen}$ and $T_{sim\_spe}$ are both 0.9. The selected candidate solutions were included within the red rectangle and the selected final solution (highest AUC) was marked in red.

We evaluated the performance of our MCMO by comparing it to the iterative multi-objective immune algorithm (IMIA), which adopts the traditional sensitivity and specificity as the optimized objective functions (Zhou *et al*, 2017) (figure 8). The two methods were compared with the unpaired *T* test at a significance level 0.05 (table 6). Results were similar based on AUC, accuracy, sensitivity and specificity (*P*-value> 0.05). For *VHL* and *PBRM1* genes, SMO achieved a little higher AUCs. For all three genes, SMO achieved significantly higher similarity scores (*P*-value <= 0.01), indicating that these results are more reliable. For the prediction result with higher AUC, the difference of $f_{sim\_sen}$ or $f_{sim\_spe}$ between SMO and IMIA is small. For example, for the *BAP1* gene, the difference of $f_{sim\_sen}$ and that of $f_{sim\_spe}$ are 0.03 and 0.02. However, for the prediction result with lower AUC, such as for the *PBRM1* gene, the difference of $f_{sim\_sen}$ is 0.13 and that of $f_{sim\_spe}$ is 0.07.

Table 5. Results of different prediction models (MOMC vs. single classifier)

| Gene | Classifier | AUC | Accuracy | Sensitivity | Specificity |
|---|---|---|---|---|---|
| *VHL* | SVM | 0.71±0.06 | 0.69±0.05 | **0.88±0.07** | 0.37±0.11 |
| | NB | 0.33±0.14 | 0.67±0.05 | 0.84±0.10 | 0.36±0.16 |
| | LR | 0.73±0.04 | 0.71±0.05 | 0.73±0.05 | 0.67±0.09 |
| | KNN | 0.80±0.04 | 0.78±0.03 | 0.84±0.04 | 0.66±0.08 |
| | DT | 0.67±0.06 | 0.68±0.07 | 0.71±0.08 | 0.61±0.11 |
| | DA | 0.71±0.05 | 0.66±0.05 | 0.68±0.11 | 0.64±0.10 |
| | MCMO | **0.88±0.01** | **0.81±0.02** | 0.79±0.04 | **0.86±0.02** |
| *PBRM1* | SVM | 0.68±0.06 | 0.63±0.06 | 0.52±0.14 | 0.70±0.06 |
| | NB | 0.41±0.09 | 0.58±0.05 | 0.52±0.11 | 0.62±0.08 |
| | LR | 0.62±0.10 | 0.63±0.09 | 0.52±0.14 | 0.69±0.09 |
| | KNN | 0.67±0.10 | 0.64±0.10 | 0.65±0.16 | 0.62±0.11 |
| | DT | 0.52±0.08 | 0.54±0.07 | 0.44±0.10 | 0.60±0.10 |
| | DA | 0.59±0.07 | 0.59±0.06 | 0.54±0.10 | 0.62±0.09 |
| | MCMO | **0.86±0.02** | **0.78±0.02** | **0.75±0.02** | **0.80±0.02** |
| *BAP1* | SVM | 0.80±0.07 | 0.81±0.03 | 0.57±0.11 | 0.85±0.04 |
| | NB | 0.74±0.11 | 0.02±0.02 | 0.24±0.0 | 0.88±0.02 |
| | LR | 0.69±0.08 | 0.81±0.03 | 0.54±0.21 | 0.84±0.04 |
| | KNN | 0.72±0.06 | 0.80±0.05 | 0.57±0.10 | 0.83±0.06 |
| | DT | 0.54±0.09 | 0.81±0.06 | 0.24±0.18 | 0.89±0.08 |
| | DA | 0.82±0.08 | 0.80±0.05 | 0.63±0.20 | 0.82±0.04 |
| | MCMO | **0.93±0.02** | **0.90±0.02** | **0.87±0.02** | **0.90±0.03** |

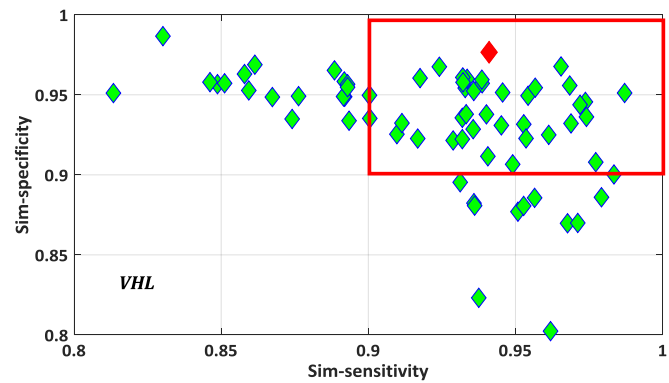

(a)

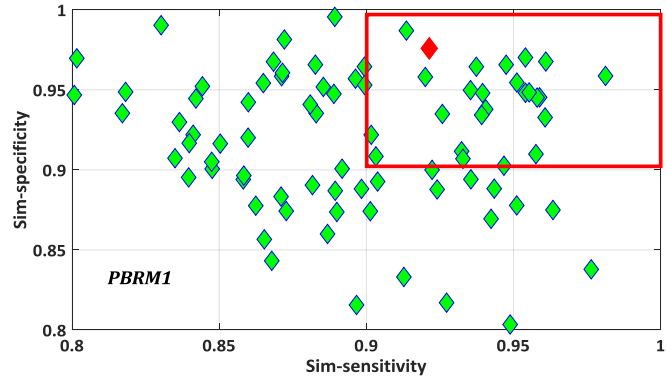

(b)

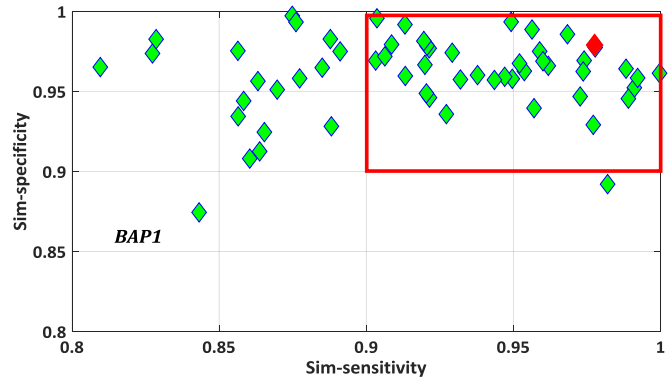

(c)

**Figure 7.** Pareto-optimal solution set (green rhombus), the green rhombus within the red rectangle are candidate solutions which satisfy $f_{sim\_sen} > T_{sim\_sen}$ and $f_{sim\_spe} > T_{sim\_spe}$. $T_{sim\_sen} = T_{sim\_spe} = 0.9$. The best solution selected for SMO is indicated by the red rhombus. (a) *VHL*; (b) *PBRM1*; (c) *BAP1*.

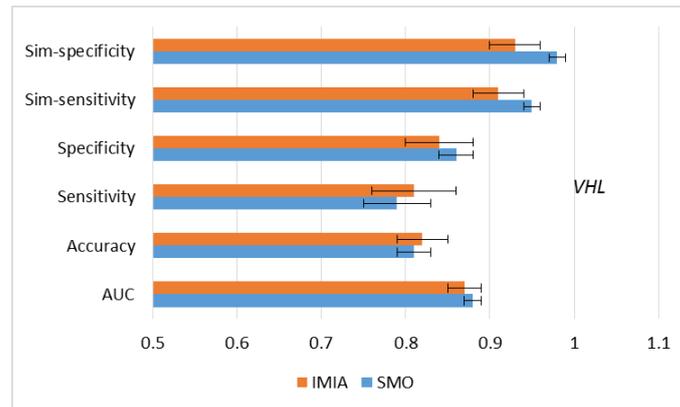

(a)

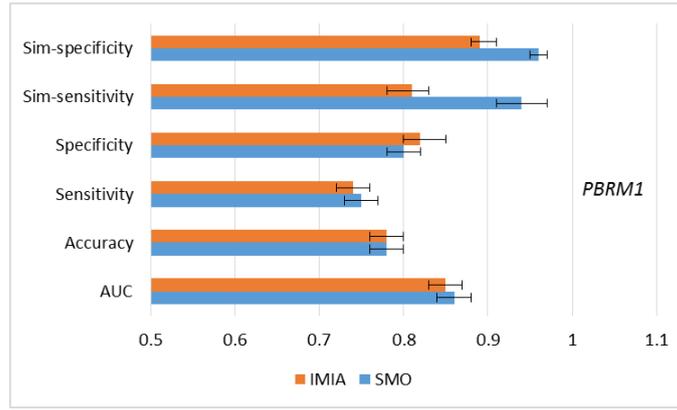

(b)

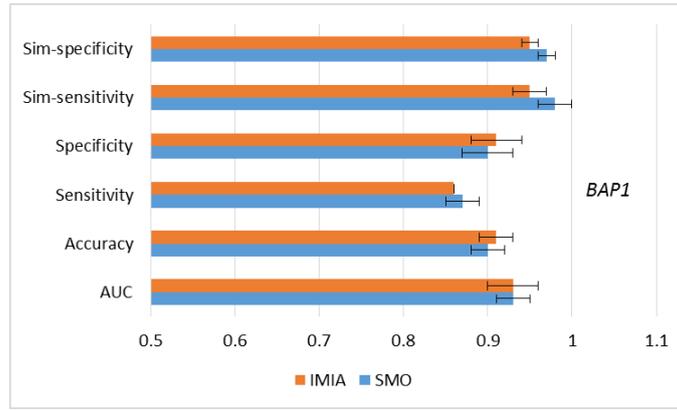

(c)

**Figure 8 .** Results of using different objective functions. (a) *VHL*; (b) *PBRM1*; (c) *BAP1*.

**Table 6. Results of *P*-values compared between SMO and IMIA**

| Gene | AUC | Accuracy | Sensitivity | Specificity | Sim-sensitivity | Sim-specificity |
|---|---|---|---|---|---|---|
| *VHL* | 0.116 | 0.449 | 0.226 | 0.206 | **0.003** | **0.0002** |
| *PBRM1* | 0.412 | 0.690 | 0.355 | 0.355 | **<0.0001** | **<0.0001** |
| *BAP1* | 0.574 | 0.536 | 0.330 | 0.472 | **0.01** | **0.007** |

### *3.5 Comparative study of fusing method*

We used the ER approach (equation (4)) for fusing the output of different classifiers. The classic weighted fusion (WF) method is used for comparison, as:

$$P = \sum_{i=1}^{M} P_i w_i \qquad (17)$$

where $P_i$ is the individual classifier output probability and $w_i$ is the relative weight. SMO is used in both fusion strategies, and the comparative results are shown in figure 9. The two methods were compared with the unpaired *t* test at a significance level 0.05 (table 7). For *VHL* and *PBRM1* genes, the ER approach achieved higher AUCs (*P*-value < 0.05). Also, sim-sensitivity and sim-specificity in ER are higher than WF (*P*-value <= 0.02), which indicates that more reliable results can be obtained when using ER fusion.

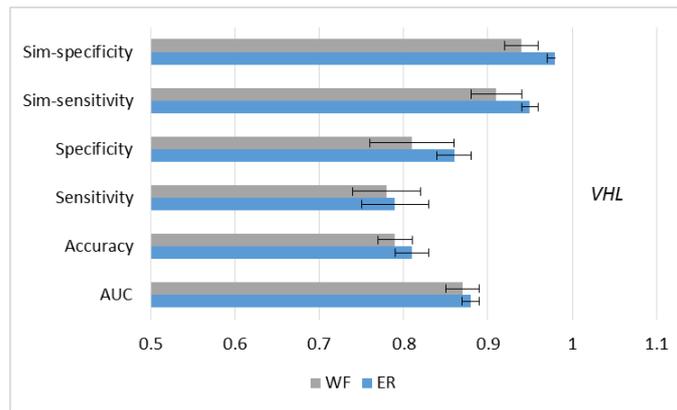

(a)

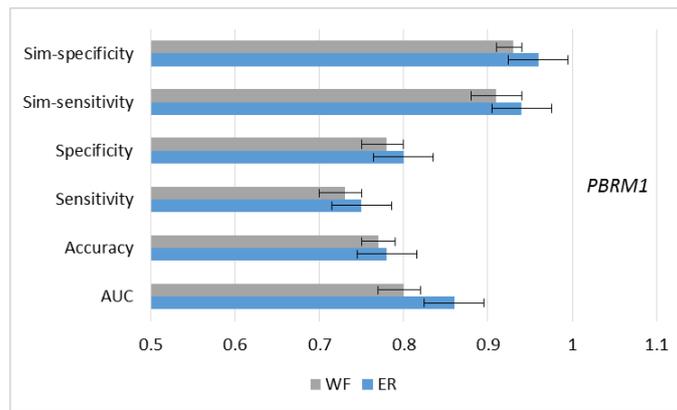

(b)

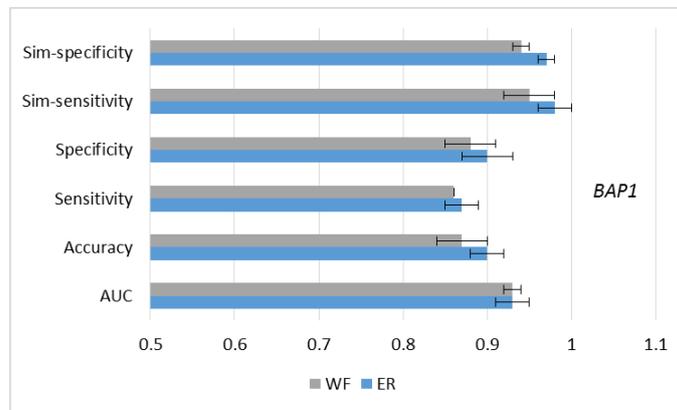

(c)

**Figure 9.** Results of using different fusion strategies (ER vs. WF). (a) *VHL*; (b) *PBRM1*; (c) *BAP1*.

**Table 7. Results of *P*-values compared between ER and WF**

| Gene | AUC | Accuracy | Sensitivity | Specificity | Sim-sensitivity | Sim-specificity |
|---|---|---|---|---|---|---|
| *VHL* | **0.021** | 0.083 | 0.760 | **0.018** | **0.007** | **<0.0001** |
| *PBRM1* | **0.0001** | 0.244 | 0.306 | 0.492 | **0.04** | **0.02** |
| *BAP1* | 0.882 | 0.045 | 0.330 | 0.074 | **0.02** | 0.0009 |

## 4. Discussion

The study of the association between diagnostic imaging features and mutations is a first critical step in the radiogenomics of ccRCC (Kuo and Yamamoto, 2011). While the genomic landscape of ccRCC is first characterized by the loss of *VHL* function, recent advances in cancer genome sequencing have identified additional, prognostically significant mutations. Two hypothesis-generating studies indicated the potential association between individual CT features and mutations of the *VHL* gene, and also mutations in the *PBRM1*, *BAP1*, SETD2 and, KDM5C genes (Karlo *et al*, 2014; Shinagare *et al*, 2015). The CT features identified by the radiologists were primarily morphological (e.g. necrosis, ill or well defined margins, renal vein invasion). In contrast, all of our features were quantitative descriptors extracted from the contoured tumor image. The feature extraction was automated, eliminating subjectivity and improving reproducibility.

The most frequently selected features varied depending on the gene. The most frequently selected features of the *VHL* gene were intensity features (Mean and Kurtosis), which described the mean values and intensity distribution in tumor volume. Intensity and texture features were found to be important in the *PBRM1* predictive model. The most selected features of *PBRM1* consisted of five intensity features that measured the intensity distribution in tumor volume, and four texture features that measured the local differences within an image. Karlo *et al* (2014) reported that nodular, heterogeneous enhancement and visibility of intratumoral blood vessels in tumors were more common among ccRCCs with underlying *VHL* mutations. Also, investigators found an association between a well-defined tumor margin and the *VHL* mutation, and an association between solid ccRCC and mutations in *VHL* and PBRM1. However, Shinagare *et al* (2015) did not observe any imaging characteristics associated with *PBRM1* and *VHL* mutations. Because both evaluations were subjective and their features were morphological, we were unable to directly compare our quantitative results with those morphological features.

The most frequently selected features of *BAP1* consisted of one geometry feature, one intensity feature, and four texture features. Texture features were the most prominent contributors in the *BAP1* predictive model. In a study by Shinagare *et al* (2015), the *BAP1* mutation was found to be associated with ill-defined tumor margins and calcification. We did not study tumor margins because they are not a quantitative feature. However, the presence of calcification may be associated with selected features such as Homogeneity and Variance.

We proposed a MCMO radiogenomics model that predicts gene mutations in ccRCC. Multi-classifier models can fully use information extracted by different classifiers and potentially improve prediction accuracy. We used all six different classifiers without considering the performance of individual classifiers. In future work, we will test the performance of single classifiers, and remove those with lower performance for the multi-classifier model. SVM and NB results for the *VHL* gene were found to be poor (table 5), therefore eliminating them and using LR, KNN, DT and DA for fusion may improve prediction results.

Both similarity-based sensitivity and specificity were considered simultaneously as the objects that guided construction of the predictive model. For the first time, we propose reliable outcome prediction, which refers to maximizing the similarity of output probability and true label (probability is 1). Higher similarity means higher reliability. We designed similarity-based sensitivity and specificity as optimized objective functions, which differ from traditional ones. Moreover, an SMO algorithm was developed to train the model to increase accuracy and confidence of predictive results. Compared with our previous IMIA method, similarity was adopted as a non-dominated sorting criterion upon updating the solution set. Also, the solution with higher similarity was kept. These findings indicate that the prediction results are more reliable (higher $f_{sim\_sen}$ and $f_{sim\_spe}$) when using our model. As for the fusion strategy, the ER approach is better than the classic WF in terms of similarity-based optimization. SMO algorithm and ER approach both contribute to reliability increase.

Our study presents a number of limitations. First, the number of patients used in our study is relatively small. In our model, eight parameters (2 parameters of SVM and 6 weights) need to be estimated and two-fold cross-validation were used. While two-fold cross-validation results showed that our model achieved satisfactory results, a larger dataset and multi-classifier fusion could help to reduce the potential risk of overfitting (Dietterich, 2000). In a future work, we can apply this MCMO model in different classification where dataset of larger size is available. Second, recent advances in genetics have led to the identification of several mutations associated with ccRCC, including those involving the *VHL*, *BAP1*, *PBRM1*, SETD2, MUC4 and KDM5C genes. The genomic information of all six genes were available for the 24 patients in TCGA/TCIA data collection. However, genomic information of the three genes (SETD2, MUC4 and KDM5C) was not available for most of the 33 UTSW cases used in the present study. Thus, we only considered *VHL*, *PBRM1* and *BAP1* genes in this work. Third, the feature stability was not addressed in the current study. Our patient data was acquired by different CT scanners at different institutions with different protocols, which resulted in differences in pixel size and slice thickness, while the differences of scanner and protocol have influence on feature calculations (Mackin *et al*, 2015). Additionally, tumor delineation was conducted by one physician and reviewed by another physician in this study. This could also introduce inter-observer variability in tumor delineation. Standardization of image acquisition protocols, automatic segmentation or consensus contours from more physicians may further improve the performance of the model developed in this work.

## 5. Conclusion

We proposed a multi-classifier multi-objective (MCMO) radiogenomics model that predicts *VHL*, *PBRM1*, and *BAP1* gene mutations in ccRCC using quantitative CT feature set. Using our feature selection strategy and model, we achieved a predictive AUC greater than 0.85 for all three genes. Compared to single classifiers, multi-classifiers fused through ER and trained by developed SMO algorithm can greatly improve prediction accuracy and reliability. In MCMO, the concept of reliable outcome prediction was first proposed and applied to the optimization procedure, generating more reliable results. The MCMO model should not only be applied to radiogenomics, but also to solving other outcome prediction problems in medicine.

**Acknowledgement**
This work was supported in part by the American Cancer Society ACS-IRG-02-196 (J. Wang), the US National Institutes of Health P50CA196516 (J. Brugarolas, P. Kapur, I. Pedrosa) and R01CA154475 (I. Pedrosa), and the National Natural Science Foundation of China 61401349 (X. Chen) and 61571359 (X. Mou). The authors thank Dr. Damiana Chiavolini for editing the manuscript.